\documentclass[aps,twocolumn,epsf]{revtex4} 
\usepackage{graphicx}
 
\begin{document}
   
 

\title{Mixing-demixing  
in a trapped degenerate 
fermion-fermion mixture}
 
\author{Sadhan K. Adhikari\footnote{Electronic
address: adhikari@ift.unesp.br\\
URL: http://www.ift.unesp.br/users/adhikari/index.html}}
\affiliation
{Instituto de F\'{\i}sica Te\'orica, UNESP $-$ S\~ao Paulo State
University,  01.405-900 S\~ao Paulo, S\~ao Paulo, Brazil}

\date{\today}
 
 
\begin{abstract}

We  use a time-dependent dynamical mean-field-hydrodynamic
model to study  mixing-demixing 
in a degenerate fermion-fermion 
mixture (DFFM). 
It is demonstrated that with the increase of  interspecies repulsion 
and/or trapping frequencies, 
a
mixed state of DFFM could turn into a fully demixed state in  both
three-dimensional 
spherically-symmetric as well as quasi-one-dimensional configurations. 
Such a demixed state of a DFFM could be experimentally realized by 
varying 
an external magnetic field near  
a  fermion-fermion
Feshbach resonance,  which will result in an increase of
interspecies fermion-fermion repulsion, and/or by increasing the external
trap frequencies.

\end{abstract}

\maketitle

\section{Introduction} 
 
Recent successful observation of degenerate  boson-fermion mixture
(DBFM) of
trapped alkali-metal atoms by different experimental groups
\cite{exp1,exp2,exp3,exp4} initiated intensive experimental
studies of different novel phenomena \cite{exp5,exp5x,exp6}. 
Among these
experiments, apart from the study of DBFM
$^{6,7}$Li \cite{exp3}, $^{23}$Na-$^6$Li
\cite{exp4} and $^{87}$Rb-$^{40}$K \cite{exp5,exp5x,bongs},  
there have been studies of degenerate fermion-fermion mixtures (DFFM)
of two components of
$^{40}$K \cite{exp1} and $^6$Li \cite{exp2,exp6}. 
The collapse in a DBFM of 
$^{87}$Rb-$^{40}$K atoms has been observed \cite{exp5} and studied
\cite{zzz} by
Modugno {\it et
al.} and by  the present author \cite{ska}  and more recently by Ospelkaus
{\it
et
al.}
\cite{bongs}. There have also been investigations 
on   the formation of 
bright \cite{fbs2} and dark \cite{fds} solitons  
in a  DBFM.

Several theoretical investigations  \cite{md1,md2,yyy1} of a trapped
DBFM
studied  the 
phenomenon of mixing-demixing when the boson-fermion repulsion is
increased. 
For a weak boson-fermion repulsion at 0 K both the Bose-Einstein
condensate
(BEC) and the degenerate Fermi gas (DFG) 
have  maxima of  probability density 
at the center of
the harmonic trap. However, with the increase of  boson-fermion
repulsion, the maximum of the probability density of the 
DFG could be  slowly expelled
from the central
region. With further increase of  boson-fermion
repulsion, the DFG could be  completely  expelled from the central region 
which will
house
only the BEC. This phenomenon has been termed mixing-demixing in a
DBFM.   The phenomenon of demixing has drawn some attention lately as 
in a demixed state 
an exotic configuration of the
mixture is formed, where there is practically no overlap between the two
components
and one can be observed and studied independent of the other.
It has been argued \cite{yyy1} that
such a demixed state in DBFM
should be possible experimentally by increasing the
interspecies scattering length 
near a 
Feshbach
resonance. 
In view of this it is of interest to see if, in addition to a DBFM, such a
demixed state appears in
a DFFM.

The purpose of this paper is to study and
illustrate 
this
mixing-demixing phenomenon in a  trapped DFFM 
using a
coupled time-dependent
mean-field-hydrodynamic model inspired by the
success of a similar model  in the
investigation of fermionic collapse \cite{ska} and bright
\cite{fbs2} and dark \cite{fds} solitons in a DBFM.
The conclusions of the study on 
bright soliton \cite{fbs2} are in
agreement with a microscopic study \cite{bongs1}, and 
those 
on collapse \cite{ska}  are
in agreement with experiments \cite{exp5,bongs}. 
There have been prior suggestions of mixing-demixing in  a  trapped 
DFFM upon an increase of interspecies repulsion. Roth and Feldmeier
\cite{demix1} discuss the possibility of 
demixing mathematically using a energy density functional approach.
Amoruso {\it et al.}  \cite{demix2} demonstrate demixing numerically using
model
hydrodynamic flow equations. In contrast to these previous
time-independent studies for stationary states, the present study relies
on a time-dependent formulation and demonstrates demixing also for  an
increase
of the trapping frequencies and should be considered complimentary to
these previous investigations. The present suggestion of demixing with an
increase of trapping frequency is easier to implement
experimentally than increasing the interspecies scattering length using a 
Feshbach resonance.

We  look for mixing-demixing  in 
a trapped DFFM  in the
spherically-symmetric as well as quasi-one-dimensional configurations. For
a weak fermion-fermion repulsion, we
find the formation of a fully mixed state with both the fermion
clouds occupying the central region of the trap. If the fermion numbers of
the two types of fermions are largely different, a fully demixed state of
DFFM appears with the increase of fermion-fermion repulsion, 
when the component with 
smaller number of fermions is completely expelled from the
central region of the trap which is occupied only by the 
component with the larger
number of fermions.  This demixing also appears for increasing trap
frequencies which increases the fermion density and hence fermion-fermion
repulsion  thus resulting in a demixing. 
To the best of our knowledge this is the first study of
mixing-demixing in a DFFM using a time-dependent formulation capable of
investigating nonequilibrium states. Previous studies of demixing
\cite{demix1,demix2}
employed 
time-independent formulations appropriate for stationary states only.

In Sec. II we present the mean-field-hydrodynamic model we use in our
investigation in the spherically-symmetric as well as
quasi-one-dimensional configuration. The
numerical results are presented in Sec. III where we illustrate
demixing with the increase of interspecies repulsion and of the
harmonic trapping potential. Finally, in Sec. III a brief summary of our
findings are given.

\section{Mean-field-hydrodynamic model} 

A proper description of a DFFM should employ
\cite{yyy1,yyy}
a fully antisymmetrized microscopic
Slater determinant wave function for each
component. However, a simplified mean-field-hydrodynamic Lagrangian for
a DFG has been used successfully to study  a DBFM
\cite{fbs2,fds}, 
which we shall use in the present investigation. The virtue of the
mean-field model over a  microscopic description is its simplicity and
good predictive power.
To  develop a set of practical time-dependent
mean-field-hydrodynamic
equations for a DFFM, we consider   the
following Lagrangian density \cite{fbs2,fds}
\begin{eqnarray}\label{yy}
&{\cal L}& = \frac{i}{2}\hbar \sum_{j=1,2}\left(
\Psi_j \frac{\partial \Psi_j \- ^*}{\partial t} - \Psi_j \- ^  
*
\frac{\partial \Psi_j
}{\partial
t} 
\right)+ g_{12}  n_1n_2 \nonumber \\ 
&+& \sum_{j=1}^2 \left(\frac{\hbar^2 |\nabla
  \Psi_j |^2 }{6m_j}+
V_j n_j+\frac{3}{5}  A_j  n_j ^{5/3}\right),
\end{eqnarray}
where  $m_j$ is the
mass
of component $j (=1,2)$, $\Psi_j$ a complex probability amplitude,
$n_j=|\Psi_j|^2$ the real probability density,
$N_j \equiv  \int d{\bf r} n_j({\bf r}) $  the
number,  
$A_j$ $=\hbar^2(6\pi^2)^{2/3}/(2m_j)$, and $V_j$ is the confining 
trap.
Here the  interspecies
coupling is 
$g_{12}=2\pi \hbar^2 a_{12}
/m_R$
with the
reduced mass $m_R=m_1m_2/(m_1+m_2),$ and  $ a_{12}$
is the interspecies
scattering length.
The interaction between intra-species fermions in
spin-polarized state is highly suppressed due
to Pauli blocking  term $3A_jn_j^{5/3}/5$ 
and has been neglected in  Eq. (\ref{yy})  and will be
neglected throughout.
The kinetic energy terms in this equation  $\hbar^2|\nabla
\Psi_j|^2/(6m_j)$
are derived from a hydrodynamic equation for the
fermions
\cite{capu} and  contribute little to this problem compared to the
dominating Pauli blocking term   in  Eq. (\ref{yy}).
However, the  inclusion of the kinetic energy terms in  Eq. (\ref{yy})
leads
to a smooth  solution for the probability density everywhere
\cite{fbs2}. To keep the algebra simple and 
without losing generality in our calculation  we shall 
take
equal
fermion masses:
$m_1=m_2\equiv m/3.$ This simulates well the fermion mixtures of
ground and excited states of $^6$Li \cite{exp2} and $^{40}$K \cite{exp1}
atoms observed
experimentally.  
The Lagrangian density of each fermion component in
Eq. (\ref{yy}) is
identical to that used in Refs.  \cite{fbs2,fds}.

The mean-field dynamical equations for the system are just the 
usual  Euler-Lagrange  (EL) equations 
\begin{equation}
\frac{d}{dt}\frac{\partial {\cal L}}{\partial   \frac{\partial
\Psi_j^*}{\partial t}}+
\sum _{k=1}^3 \frac{d}{dx_k}\frac{\partial {\cal L}}{\partial
\frac{\partial \Psi_j^*}{\partial x_k}}= \frac{\partial {\cal
L}}{\partial
 \Psi_j^*},
\end{equation}
where $x_k, k=1,2,3$ are the three space components, and 
$j=1,2$ refer to the fermion components. 

With Lagrangian density
(\ref{yy}) the following  EL   equations of 
motion are
derived in a straight-forward fashion \cite{ska,fbs2}: 
\begin{eqnarray}\label{e}  \biggr[ - i\hbar\frac{\partial
}{\partial t}
-\frac{\hbar^2\nabla_{\bf r}^2}{6m_{{j}}}
+ V_{{j}}({\bf r})+A_j n_j^{2/3} 
&+& g_{{12}} n_k
 \biggr]\Psi_j=0, \nonumber \\ 
 j \ne k=1,2.  \end{eqnarray} This is essentially a time-dependent version
of a similar time-independent model used recently for fermions \cite{md2}.
For a stationary state Eqs. (\ref{e}) yield the same result as the
formulation of Ref. \cite{md2}. For a system with large number of fermions
both reduce to \cite{ska,md2} the well-known Thomas-Fermi approximation
\cite{rmp,str}:
$n_j=[(\mu_j -V_j)/A_j]^{3/2}$, with $\mu_j$ the chemical potential.

In the spherically-symmetric case we take 
 $
V_{j}({\bf
r})=\frac{1}{2}(3m_j) \omega ^2 r^2$ and 
 $\Psi_j({\bf
r};t)=\psi_j(r;t).$ 
Now  transforming to
dimensionless variables
defined by $x =\sqrt 2 r/l$,     $\tau=t \omega, $
$l\equiv \sqrt {\hbar/(m\omega)}$
and
\begin{equation}\label{wf}
\frac{ \phi_j(x;\tau)}{x} =  
\sqrt{\frac{4 \pi l^3}{N_j\sqrt 8}}\psi_j(r;t),
\end{equation}
we get \cite{ska}
\begin{eqnarray}\label{d1}
\biggr[&-& i\frac{\partial
}{\partial \tau} -\frac{\partial^2}{\partial
x^2} 
+\frac{x^2}{4} 
+ {\cal N}_{jj}
\left|\frac {\phi_j}{x}\right|^{4/3} \nonumber \\
&+&    {\cal N}_{jk}\left
|\frac {\phi_k}{x}\right|^2 
\biggr]\phi_j({ x};\tau)=0,  
\end{eqnarray}
where $j\ne k=1,2$ and 
$ {\cal N}_{jj}=3(3\pi N_j/2)^{2/3}$, 
$ {\cal N}_{jk} = 6 \sqrt 2  N_k a_{12} /l.$ 
The normalization of the wave-function components
is
given by $\int_0^\infty dx |\phi_j(x;\tau)|^2 =1, j=1,2.$

We next reduce  Eq. (\ref{e})  to a  minimal 
 quasi-one-dimensional form
in a cigar-shaped geometry  where the confining trap of anisotropy 
$\nu$ 
has the form
 $V_j({\bf r})= \frac{1}{2}3 m_j\omega^2(
\rho^2+\nu^2 z^2),$ with $\rho$ the radial vector and $z$ the axial 
vector.  
For  a cigar-shaped geometry  $\nu << 1$, 
we consider solutions of    Eq.   (\ref{e})  of the type
$\Psi_j({\bf r},t)= \sqrt N_j \varphi_j(z,t)\psi_j^{(0)}( \rho), $
where
\begin{eqnarray}
|\psi_j^{(0)}(\rho)|^2&\equiv&
{\frac{3m_j\omega}{\pi\hbar}}\exp\left(-\frac{3m_j
\omega
\rho^2}{\hbar}\right)
\end{eqnarray}
corresponds to the respective circularly symmetric
ground-state wave function in the absence of nonlinear interactions and
satisfies
 \begin{eqnarray}
-\frac{\hbar^2}{6m_j}\nabla_\rho ^2\psi_j^{(0)}
+
\frac{3}{2}m_j\omega^2\rho^2
\psi_j^{(0)}&=&\hbar\omega
\psi_j^{(0)}
\end{eqnarray} 
with normalization
$2\pi \int_{0}^\infty |\psi_j^{(0)}(\rho)|^2 \rho d\rho=1.$
Now the dynamics is carried by $ \varphi_j(z,t)$ and the radial dependence 
is
frozen in the ground state $\psi_j^{(0)}(\rho)$. The separation of
the variables is suggested by the structure of  Eq. (\ref{e}).

Averaging over the radial mode $\psi_j^{(0)}$,
i.e., multiplying Eq. (\ref{e}) 
by  $\psi_j^{(0)*}(\rho)$
and integrating over $\rho$, we obtain the following one-dimensional
dynamical equations \cite{fbs2,fds}:
\begin{eqnarray}\label{j}
 \biggr[&-& i\hbar\frac{\partial
}{\partial t}
-\frac{\hbar^2}{6m_j}\frac{\partial^2}{\partial z^2}  
+ \frac{3}{2}m_j\nu ^2\omega^2 z^2 \nonumber \\
&+& F_{jj}N_j^{2/3}|
\varphi_j|^{4/3}
+ F_{jk}N_k| \varphi_k|^2
 \biggr] \varphi_{{j}}(z,t)=0,
\end{eqnarray}
where $j\ne k= 1,2$ and 
\begin{eqnarray}
F_{jk}=g_{12}
\frac{\int_0^\infty|\psi_k^{(0)}|^2|\psi_j^{(0)}|^2\rho
d\rho}{\int_0^\infty|\psi_j^{(0)}|^2\rho d\rho}
=g_{12}{\frac{M_{12}\omega}{\pi\hbar}},
\end{eqnarray} 
\begin{eqnarray}
F_{jj}=A_j
\frac{\int_0^\infty|\psi_j^{(0)}|^{2+4/3}\rho
d\rho}{\int_0^\infty|\psi_j^{(0)}|^2\rho
d\rho} =
{\frac{3A_j}{5}}\left[
\frac{3m_j\omega}{\pi \hbar}    \right]^{2/3},
\end{eqnarray} 
where
$M_{12}= 3m_1m_2/(m_1+m_2).$
In  Eq.  (\ref{j}) 
the normalization 
is  $\int_{-\infty}^\infty |\varphi_j(z,t)|^2
dz = 1$.

In   Eq. (\ref{j}),
we introduce  the dimensionless variables
$\tau=t\nu \omega/2$,
$x=z /l_z$,
${\phi}_j=
\sqrt{l_z} \varphi_j$, with $l_z=\sqrt{\hbar/(\nu \omega m)}$,
so that
\begin{eqnarray}\label{n} 
 \biggr[ &-&  i\frac{\partial
}{\partial \tau}-\frac{\partial ^2}{\partial x^2}
+c^2 x^2     +
{\cal N}_{jk}
  \left|{{\phi}_k} \right|^2\nonumber \\
&+&
{\cal N}_{jj}
  \left|{{\phi}_j}
\right|^{4/3}
 \biggr]{\phi}_{{j}}(x,\tau)=0,\quad
 j\ne k =1,2, 
\end{eqnarray}
where
$c=1,$
${\cal N}_{jk}=12 a_{12}(N_k/\nu)/l_z,$ and
${\cal N}_{jj}=9(6\pi N_j/\nu)^{2/3}/5. $ However, by taking $c\ne 1$ we
can
modify the harmonic trap. 
In Eq. 
 (\ref{n}),
the normalization condition  is given by
$\int_{-\infty}^\infty |\phi_j(x,\tau)|^2 dx =1 $.


\begin{figure}
 
\begin{center}
\includegraphics[width=.8\linewidth]{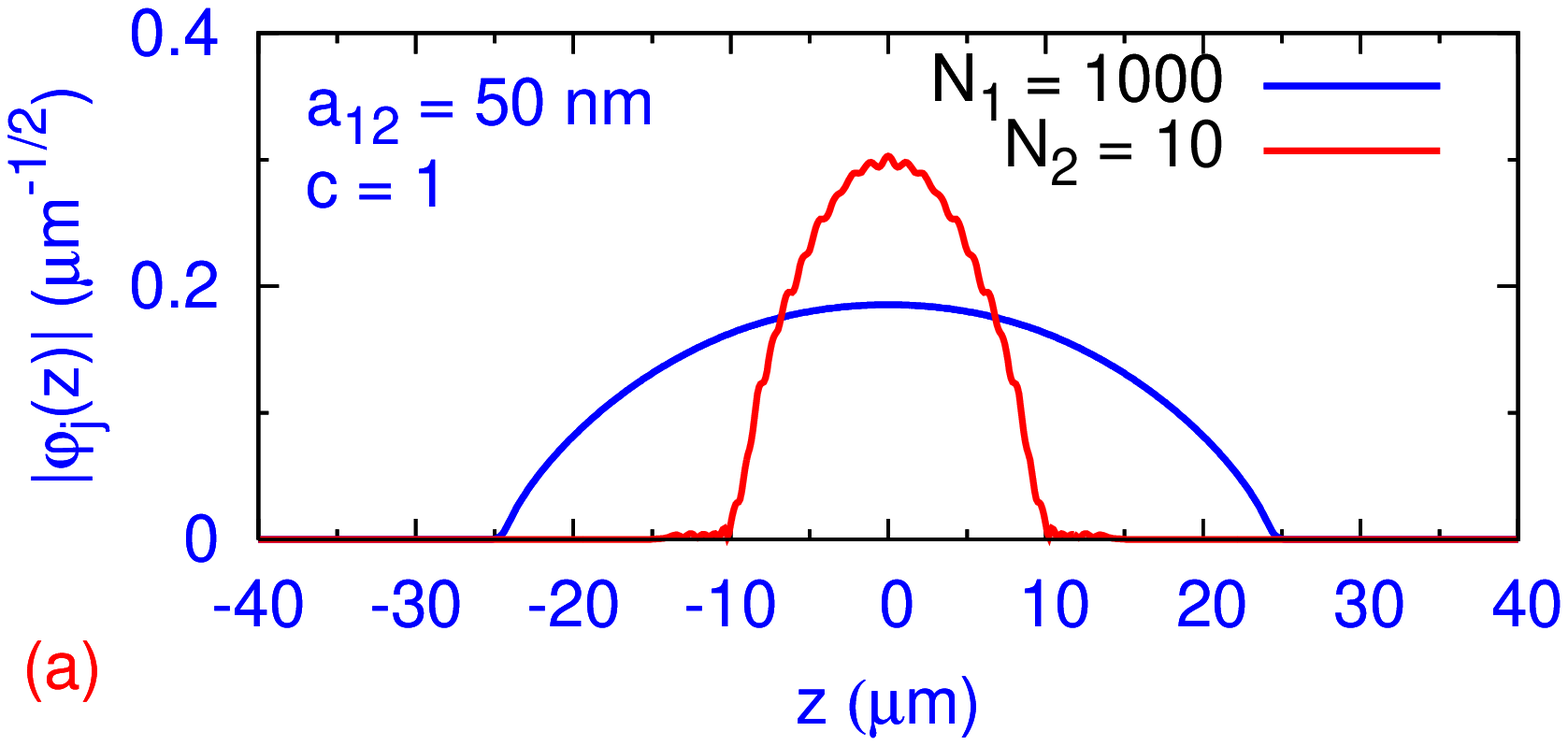}
\includegraphics[width=.8\linewidth]{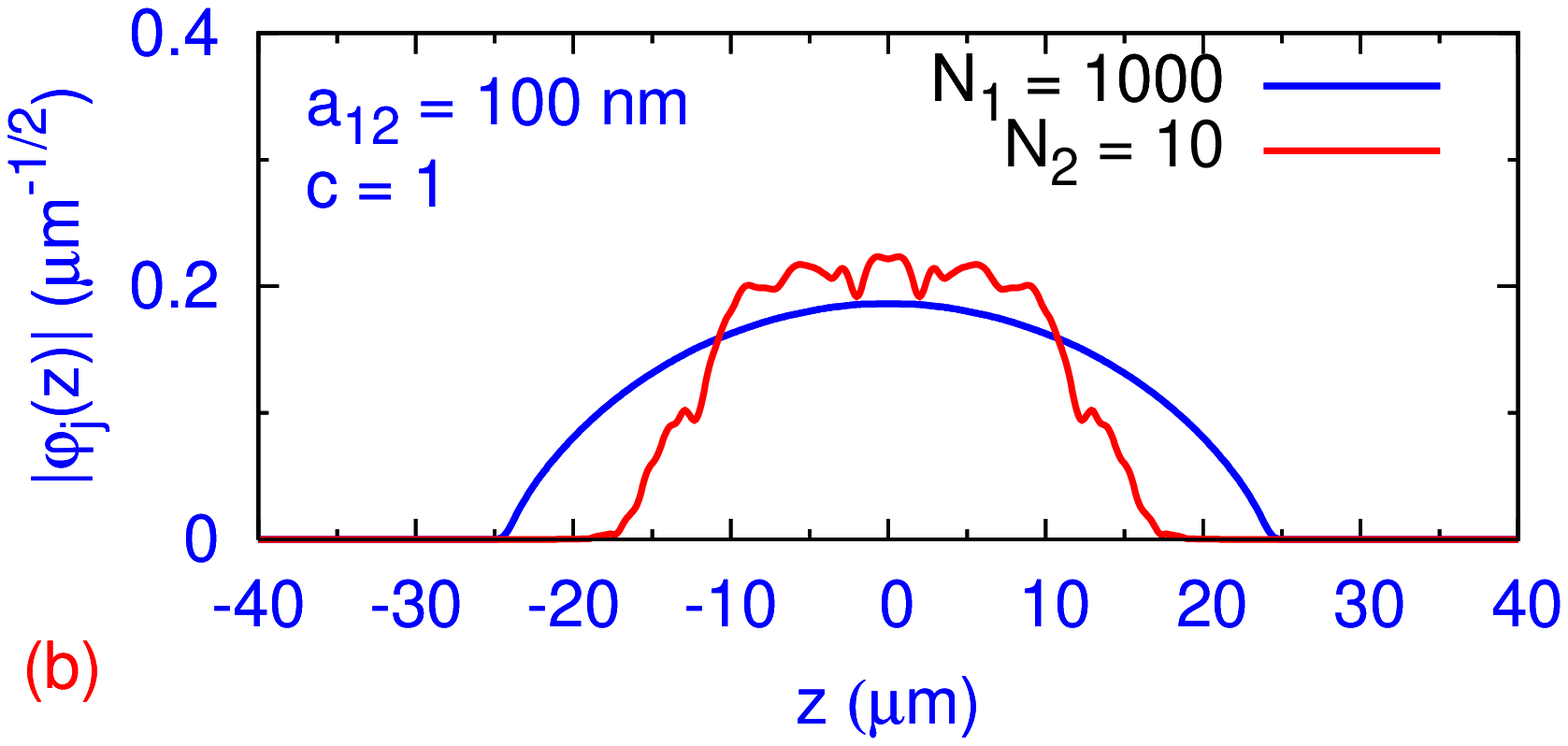}
\includegraphics[width=.8\linewidth]{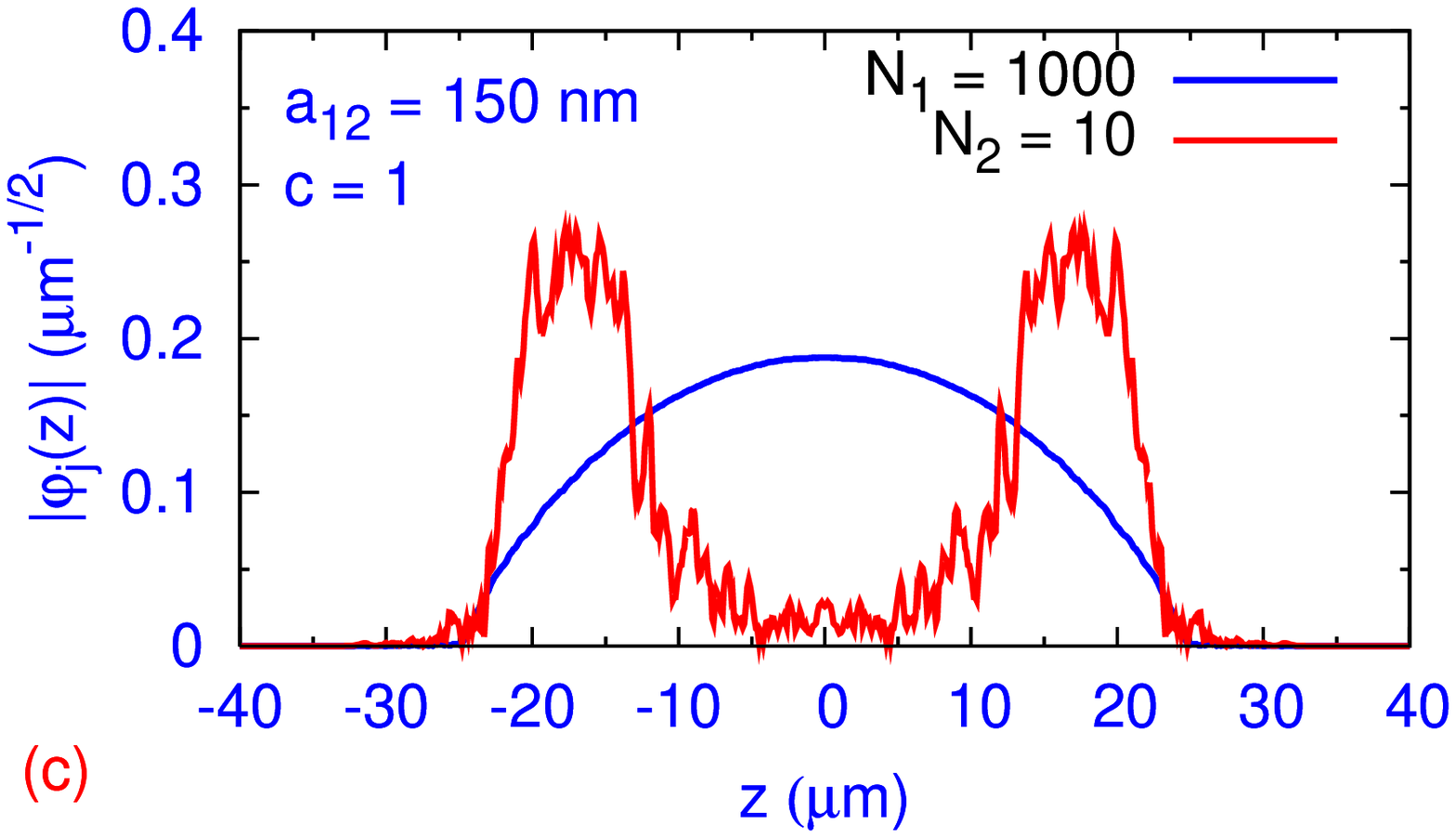}
\includegraphics[width=.8\linewidth]{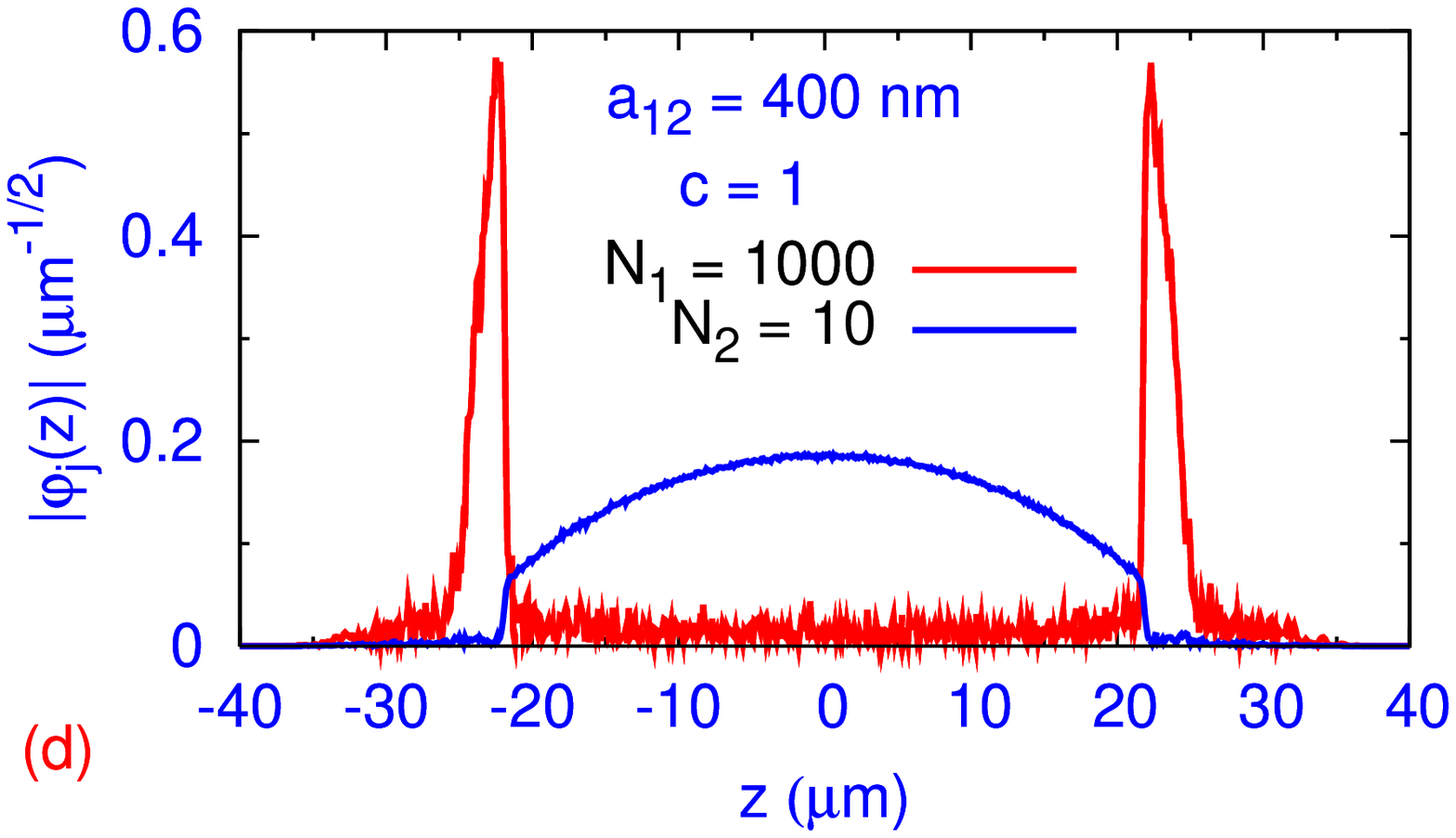}
\end{center}

\caption{(Color online) 
The fermionic profiles $\varphi_j(z)$ of Eq.  (\ref{j})
vs $z$ for $N_1=1000$ and $N_2=10$
in the quasi-one-dimensional case ($\nu = 0.1$) for $a_{12}=$ 
(a) 50 nm, (b) 100 nm, (c) 150 nm, and (d) 400 nm.
}
\end{figure}

\section{Numerical Results}
 
We solve the coupled mean-field-hydrodynamic  equations (\ref{d1}) and
(\ref{n}) numerically using a time-iteration
method based on the Crank-Nicholson discretization scheme
elaborated in  \cite{sk1}.  
We
discretize the mean-field  equations
using time step $0.001$ and space step $0.05$.
In the quasi-one-dimensional case we span $x$ from $-60$ to $60$, and in
the three-dimensional case we span $x$ from 0 to 30. 
In both cases we start with the Gaussian ground-state solution of the
linear harmonic
oscillator problem  as
input at $\tau =0$ while we set the nonlinearities ${\cal N}_{jk}={\cal
N}_{jj}= 0$.
In the three-dimensional case governed by  Eq. (\ref{d1}) we take
$\phi_j(x, \tau =0)= (\pi/2) ^{-1/4} \exp(-x^2/4)$  
and in the one-dimensional case  governed by  Eq. (\ref{n}) we take 
$\phi_j(x, \tau =0)= \pi ^{-1/4} \exp(-x^2/2)$. With these initial states
we perform time evaluation of Eqs.  (\ref{d1}) and
(\ref{n}). During the course of time evolution the nonlinear terms ${\cal
N}_{jk}$ and ${\cal
N}_{jj}$
are
slowly introduced  at the rate of 0.0001 at each time step of 0.001. The
solution evolves slowly with time and attains the final stationary
configuration after the full desired nonlinearity is introduced slowly. It
is important to introduce the nonlinearity slowly to achieve a final
stationary  configuration.   

 In our numerical investigation we take $l=1$ $\mu$m and consider the
fermions to have the mass of  $^{40}$K  atoms and to
occupy two levels of this fermionic atom. In the spherically-symmetric
case this   corresponds to  a radial
frequency $\omega \approx 2\pi \times 83$ Hz.
Consequently, the  unit of
time is  $1/\omega \approx 2$ ms, and the unit of length $l/\sqrt 2
\approx
0.7$   $\mu$m.  In the quasi-one-dimensional case we take $\nu =0.1$ and 
$l_z=1$  $\mu$m  corresponding to an axial frequency  $\omega \nu \approx
2\pi
\times 83$ Hz. Consequently, the  unit of
time is  $2/(\nu \omega) \approx 4$ ms, and the unit of length $l_z =
1$   $\mu$m.

After a small experimentation 
it is realized that both in  Eqs. (\ref{d1}) and (\ref{n}), demixing is
most
favored when $N_1$ and $N_2$ are widely different from each other. Under
this condition one of the components is much larger in size than the other
and easily expels the smaller component from the center of the trap upon
an increase in interspecies repulsion or trap frequency and we shall
consider only this asymmetric configuration here. In previous
investigations \cite{demix2} demixing in the symmetric configuration 
($N_1=N_2$) was considered.  The interspecies repulsion can be
manipulated through the scattering length $a_{12}$ by varying a background
magnetic field near a fermion-fermion
Feshbach resonance \cite{fsff}.

\begin{figure}
 
\begin{center}
\includegraphics[width=.8\linewidth]{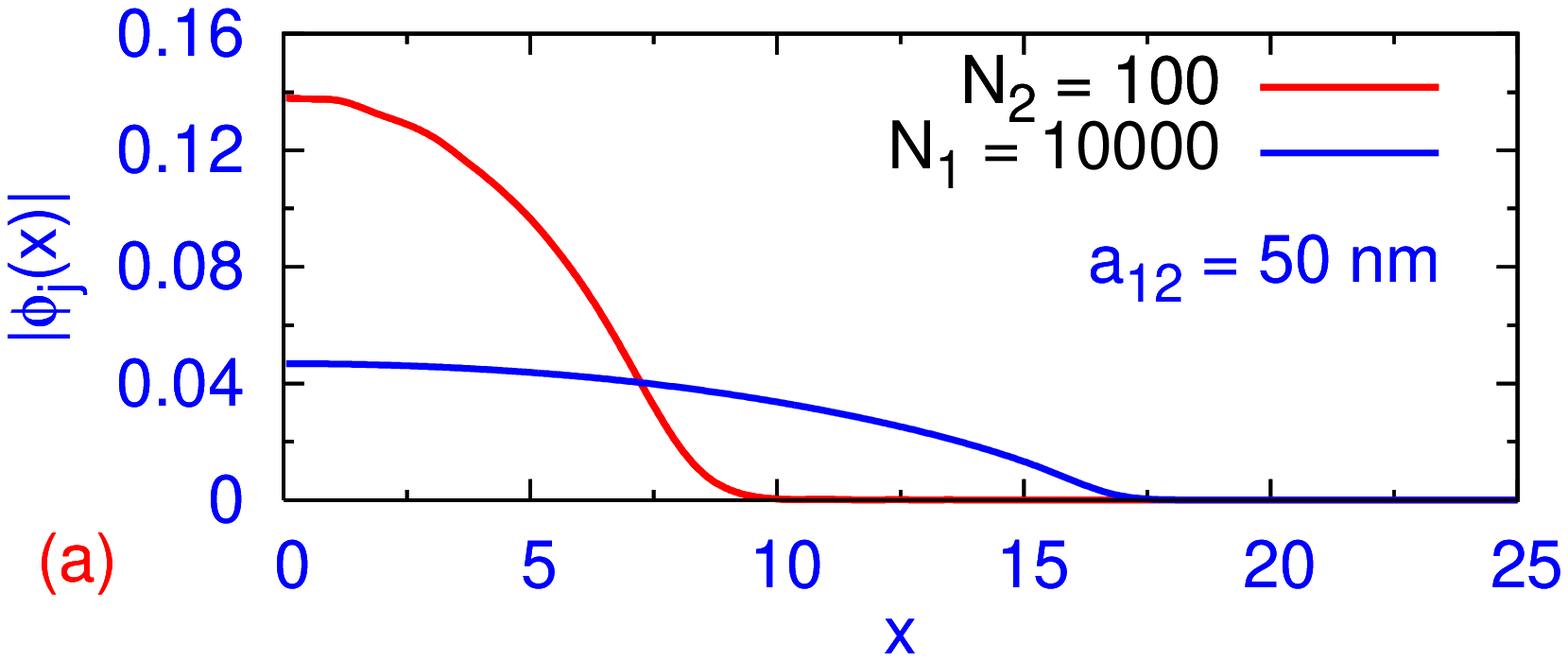}
\includegraphics[width=.8\linewidth]{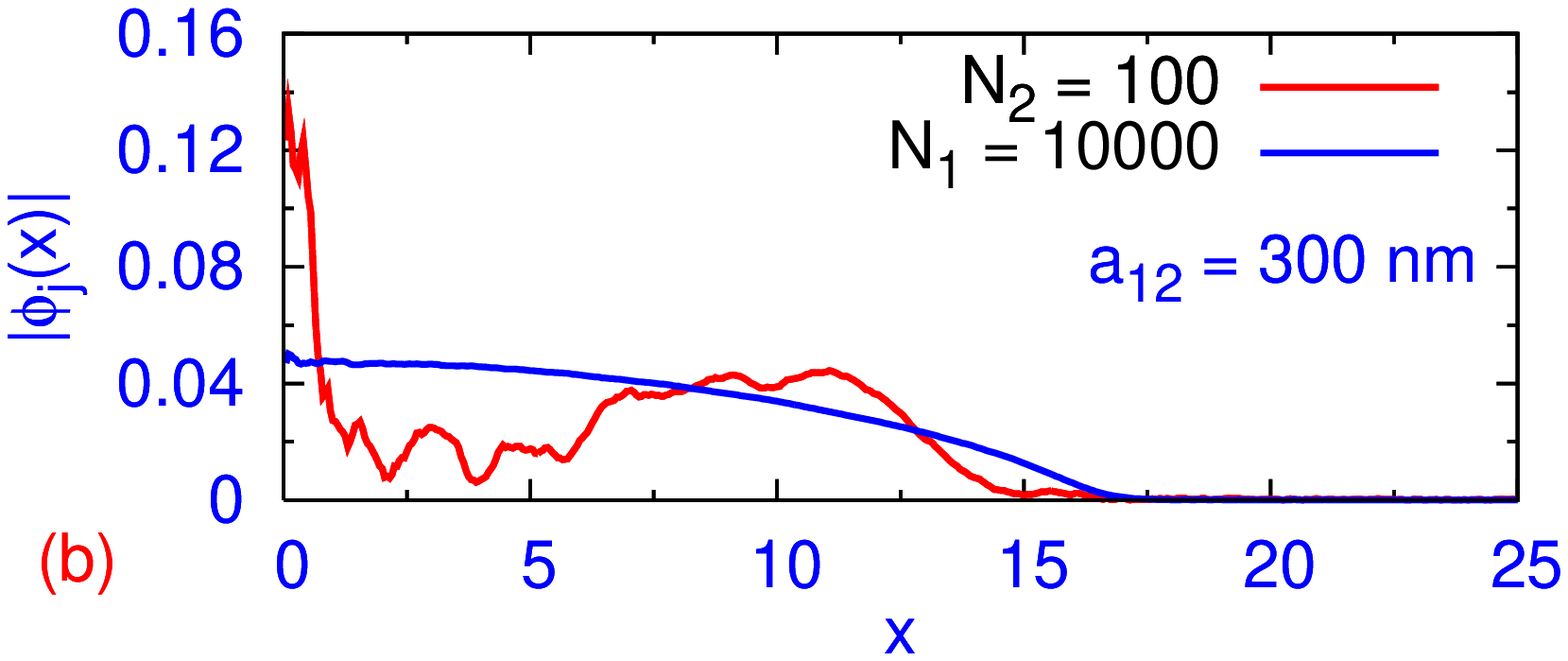}
\includegraphics[width=.8\linewidth]{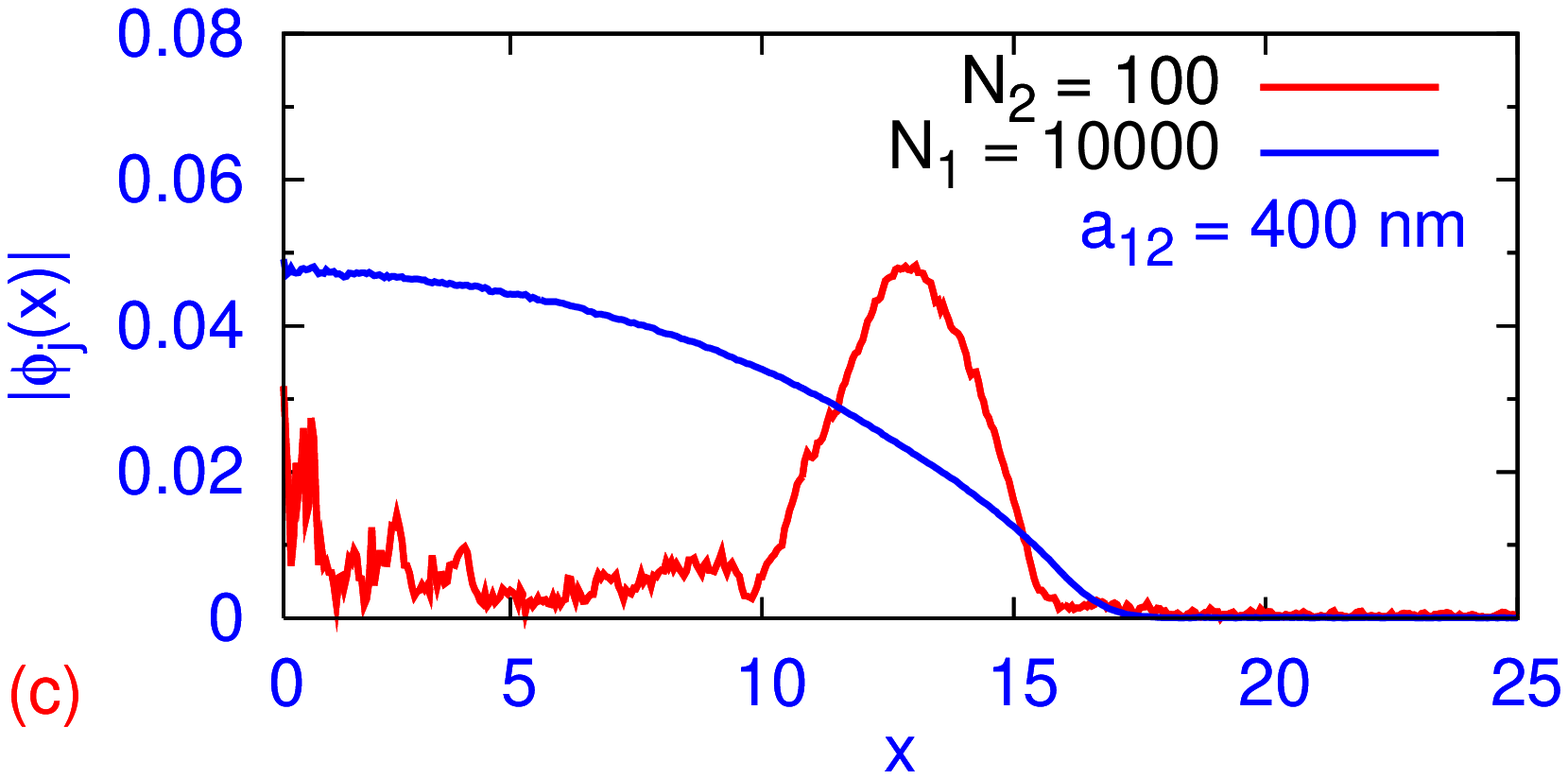}
\includegraphics[width=.8\linewidth]{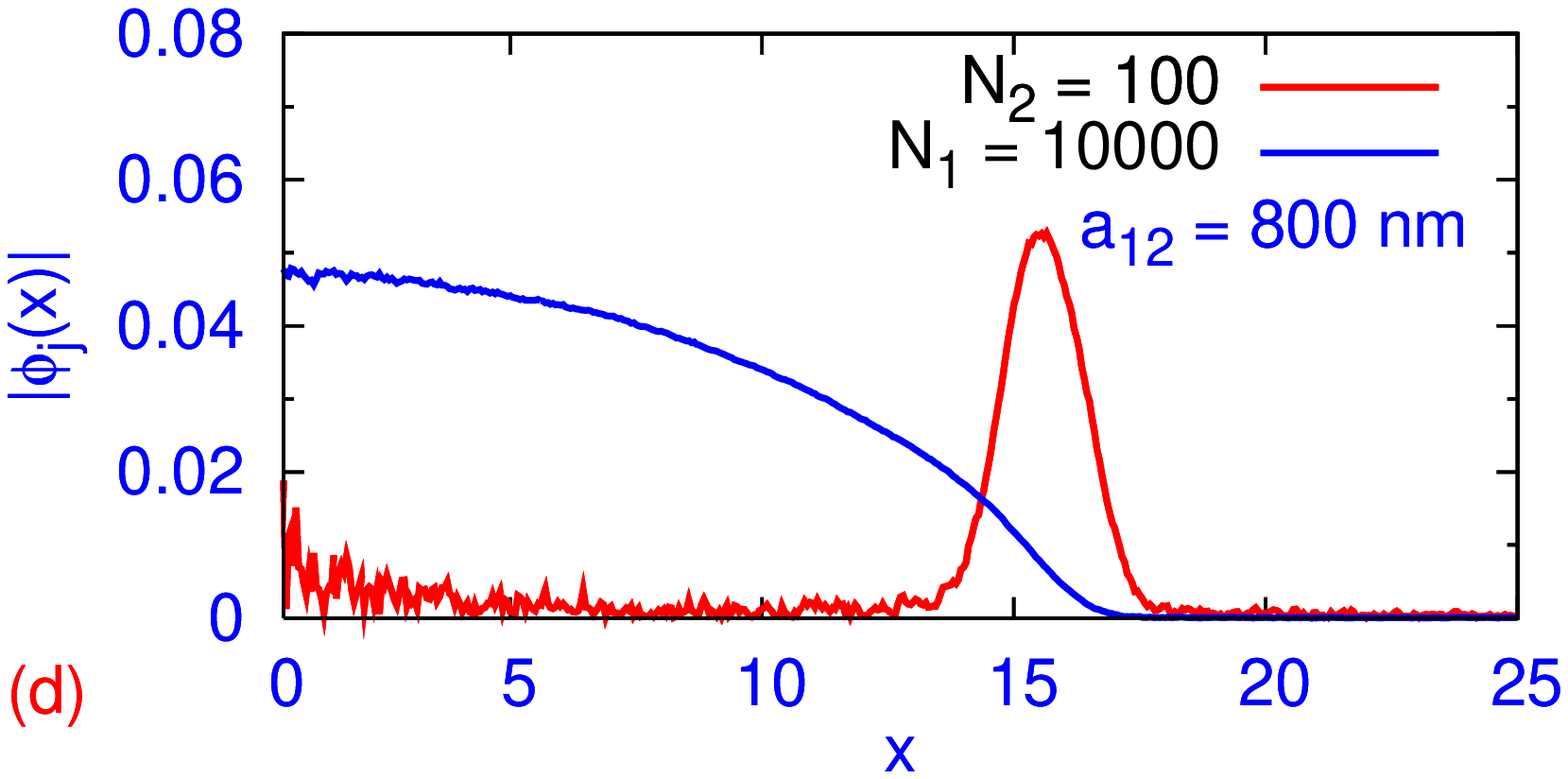}
\end{center}

\caption{(Color online) 
The fermionic profiles $\phi_j(x)$ of Eq. (\ref{d1}) vs. $x$ in
dimensionless
units for
$N_1=10000$ and $N_2=100$
in the spherically-symmetric case  for $a_{12}=$ 
(a) 50 nm, (b) 300 nm, (c) 400 nm, and (d) 800 nm.
}
\end{figure}

First, we present results for the quasi-one-dimensional case from a
solution of  Eq. (\ref{n}).
After some experimentation we take in the initial DFFM 
$N_1=1000$, $N_2=10$, and $a_{12}= 50$ nm, $\nu =0.1$. 
This
corresponds to nonlinearities ${\cal N}_{11}\approx 5917$, ${\cal
N}_{22} \approx 274$,
${\cal N}_{12}= 60$ and ${\cal N}_{21}=
6000$. The two overlapping    Gaussian-type
fermionic profiles in this case 
are shown in Fig. 1 (a). Next we increase the interspecies repulsion  by
increasing the interspecies scattering length $a_{12}$. With the increase
of repulsion
the component with smaller number of fermions 
is slowly deiven out of the central region which
continues to be 
solely occupied by the component
 with larger number of fermions. In Figs. 1 (b), (c) and (d) we
plot the results for  $a_{12}= $ 100 nm, 150 nm, and 400 nm, from where
we find that a complete demixing has occured for  $a_{12}= $ 400 nm.
In the last case two symmetrical peaks of the component with
smaller number of fermions appear
where the density of the component with larger number of fermions
is practially zero. The interesting thing to note is that in Figs.
 1 (a) $-$ (d), the profile of the degenerate gas  with larger number of
fermions 
remains practially unchanged, while the weaker component with much smaller
number of fermions is expelled from the central region with the increase
of repulsion. The mutual repulsion has practically no effect on the
``heavier"
component when it expells the ``lighter" component from the central
region.   Had we considered the symmetric case with $N_1=N_2=1000$ a much
larger $a_{12}$ would be necessary for a complete demixing.

Next we consider  demixing in the three-dimensional spherically-symmetric
case 
from the solutions of  Eq. (\ref{d1}). In this case we take in the initial
DFFM $N_1=10000$, $N_2=100,$ and $a_{12}=50$ nm. This corresponds to
nonlinearities ${\cal N}_{11}\approx 3914$, ${\cal N}_{12}\approx
42$, ${\cal N}_{21}\approx 4243$, and ${\cal N}_{22}\approx 181$.
The overlapping fermionic quasi-Gaussian
profiles of the DFGs in this case are shown in Fig. 2 (a). 
Now we increase the interspecies repulsion by increasing the scattering
length $a_{12}$. With the increase of repulsion among the two species, 
smooth Gaussian profiles of the DFFM are destroyed and 
the
component  with smaller number of fermions       
is slowly driven out of the central region which
continues to be occupied by the component with larger number of
fermions. In Figs. 2 (b), (c), and
(d) we plot the results for $a_{12}=300$ nm, 400 nm, and 800 nm,
respectively. From Figs. 2 we see that demixing has increased with the
increase of repulsion among the two components. In Fig. 2 (d) the 
component  with smaller number of fermions       
is practically located outside the region occupied by the 
component with larger number of fermions. 

\begin{figure}
 
\begin{center}
\includegraphics[width=.8\linewidth]{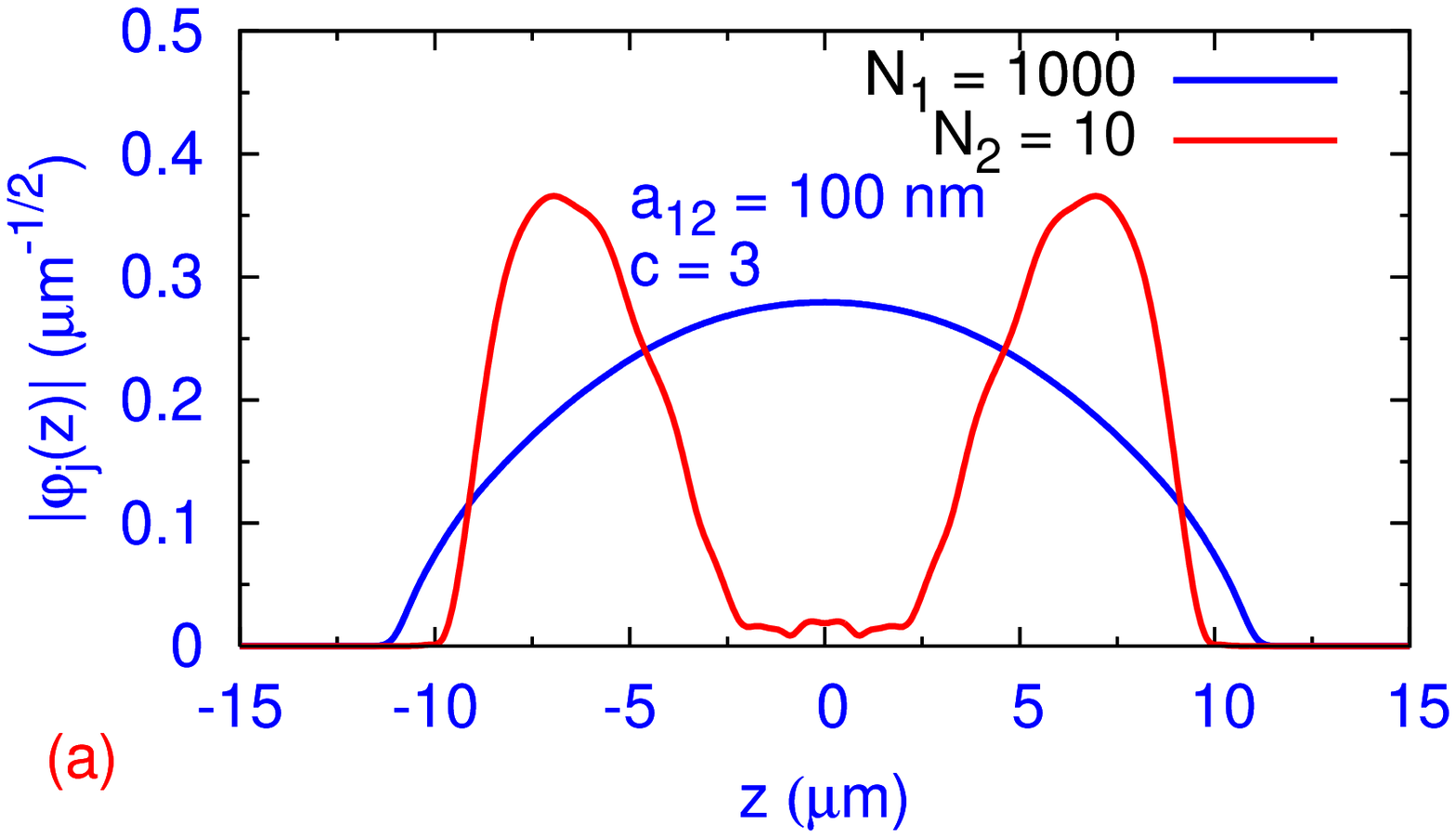}
\includegraphics[width=.8\linewidth]{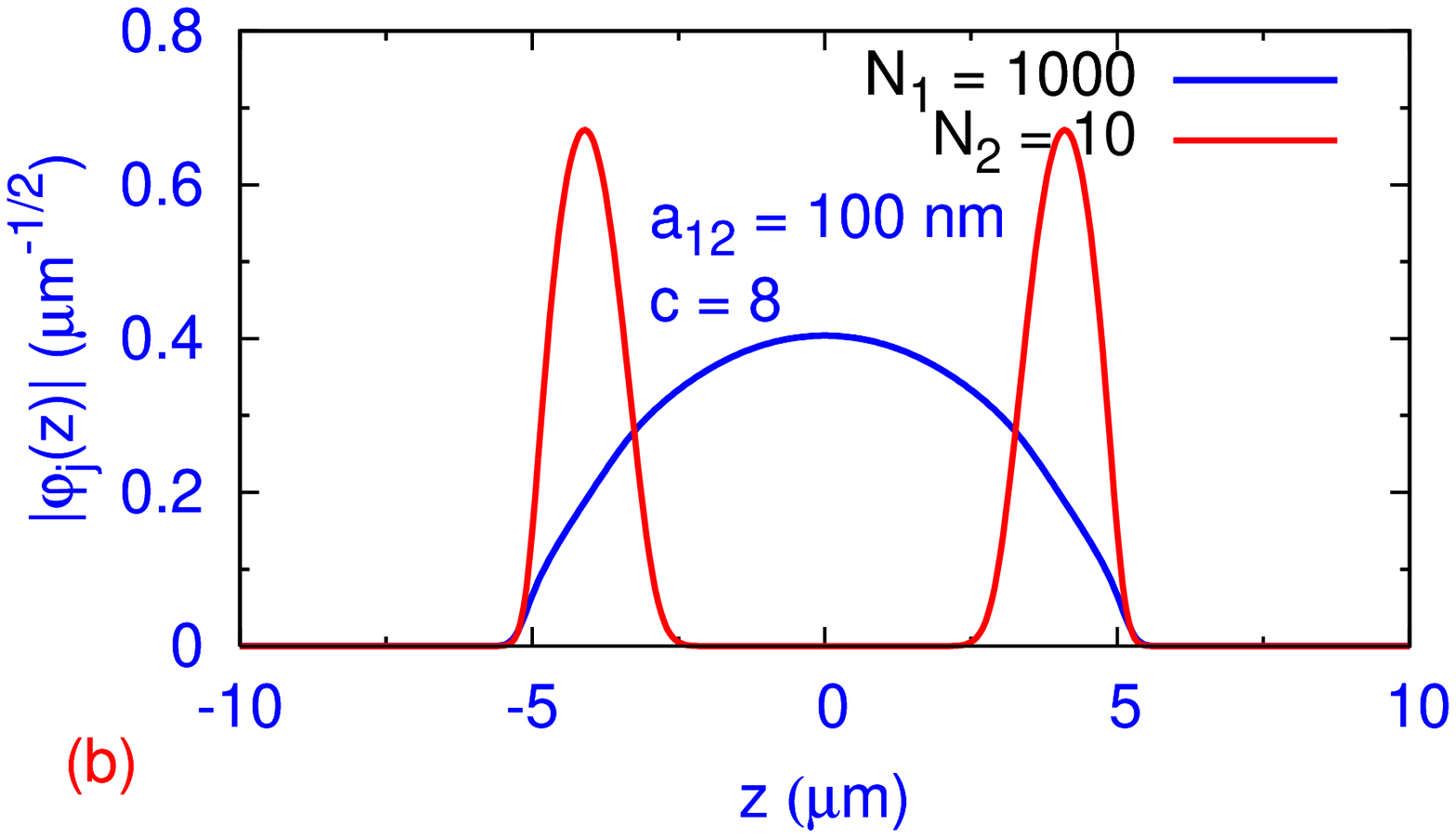}
\end{center}

\caption{(Color online) 
The fermionic profiles $\varphi_j(z)$ of Eq. (\ref{j}) vs. $z$ for
$N_1=1000$,
$N_2=10$
and  $a_{12}=100$ nm       
in the quasi-one-dimensional case ($\nu = 0.1$) for $c=$ 
(a) 3 and  (b) 8.
}

\end{figure}

The dimensionless parameter controlling the validity of the present
dilute (weakly interacting)
mean-field-hydrodynamic model is the number of particles in a
scattering
volume $|a|^3$, where $a$ is the scattering length. This can be written as
$\bar n|a|^3$, where $\bar n$ is
the average density of the gas \cite{rmp}. When  $\bar n|a|^3<<1$ the
system is said
to be dilute or weakly interacting. The opposite limit  $\bar n|a|^3>>1$
is known as the unitarity limit. It is interesting to find  out if
demixing in the three-dimensional case shown in Fig. 2 occured in the
dilute limit.  In Fig. 2 (c)  demixing is visible for $|a|= 400$ nm where
there are 10000 atoms in a sphere of radius $R\approx 18$
$\mu$m. Consequently, $\bar n= 10000/(4\pi R^3/3) \approx 0.4$
$\mu$m$^{-3}$
and $|a|^3 \approx 0.064$ $\mu$m$^{3}$, so that $\bar n|a|^3 \approx
0.025.$ So demixing starts in the dilute weakly interacting
limit and continues in the
unitarity limit.

The present condition of demixing is consistent with that
of Ref. \cite{demix2}, where they have reported demixing as a function of
the
parameter \cite{str}
\begin{equation}
\gamma \equiv \frac{E_{\mbox{int}}}{E_{\mbox{pot}}}= \alpha N^{1/6}
\frac{a}{l},
\end{equation}
with $\alpha = 2^{1/3} 3 ^{1/6}(8192/2835\pi^2)\approx 0.44$ and  $N$ the
number of atoms, $E_{\mbox{int}}$ is the interaction energy and
$E_{\mbox{pot}}$ is the potential energy of the harmonic oscillator
trap. The parameter $\gamma$ 
compares the interaction energy with the potential energy in the harmonic
trap and  provides a more reliable account of the words ``weakly
interacting". A small value of the parameter
$\gamma$ together with the condition $\bar n |a|^3 << 1$
is necessary for the DFFM to be weakly interacting.
In the situation of Fig. 2 (c) ($a=400$ nm)
demixing has started for $\gamma \approx
0.44 \times 10000^{1/6} \times 0.4
\approx 0.8$ and is absent in the situation of Fig. 2 (b)  ($a=300$ nm)
where  $\gamma \approx 0.6$.     
In the study of  Ref. \cite{demix2} demixing started at
$\gamma
\approx 0.7$, and was absent for  $\gamma \approx 0.54$
quite consistent with  present
finding. The small value of $\gamma (<1)$ obtained in the present study as
well as in Ref. \cite{demix2}
confirm the dilute or weakly
interacting nature of the DFFM in the demixing limit. 
However, according to the condition obtained in  Ref. 
\cite{demix1}, demixing should  occur only in  the unitarity limit. We do
not
know the reason for this discrepancy.
Further
investigations are needed for clarification.

Now we investigate the possibility of demixing in a DFFM by increasing
the strength of the harmonic trap. We illustrate the results only in the
quasi-one-dimensional case described by Eq. (\ref{n}). From Fig. 1 (b) we
find
that there is no real demixing in  the solution of Eq. (\ref{n})
for $N_1=1000, N_2=10$, 
$a_{12}= 100$ nm and $c=1$. 
In this case we find that demixing appears as
we increase $c$ keeping other parameters fixed. The results are
illustrated in Figs. 3 (a) and (b) for
$c=3$ and 8. The increase of parameter $c$ increases the trap frequency in
the same proportion. From Fig. 3 (a) we find that demixing already 
appears for $c=3$ and increases in Fig. 3 (b) for $c=8$.
For a  large $c$,  the appearance of  
demixing for a smaller $a_{12}$ can be understood from
the structure of  Eq. (\ref{n}). An increase of the trapping frequency
corresponds to a
decrease in the harmonic oscillator length $l_z$. Then a smaller value of
$a_{12}$ can lead to a larger nonlinearity ${\cal N}_{jk}$ as in the
definition of  ${\cal
N}_{jk}$,     $a_{12}$   is always scaled with respect to $l_z$. However,
with the increase of the trapping frequency  the scaling length $l_z$ is
reduced and the
degenerate gas is confined to a smaller region in space as can be seen
from Figs. 3 (a) and (b).

\section{Summary}

Using a coupled set of time-dependent mean-field-hydrodynamic
equations for a trapped DFFM derived using the Lagrangian density for a
DFG
suggested recently \cite{fbs2,fds}, we study the phenomenon of demixing
in a trapped DFFM.
We show that, for a repulsive 
interspecies fermion-fermion interaction, demixing is favored in a DFFM
with different number of fermions.
Demixing appears for a large value of interspecies scattering length 
$a_{12}$ denoting a large repulsion between the components. As the
interspecies repulsion increases, the fermionic component with smaller
number of fermions is expelled from the central region which is solely
occupied by the component with larger number of fermions. 
Demixing  in a trapped DFFM can be realized experimentally 
by changing  the interspecies
scattering length  $a_{12}$
to a large positive  value by exploiting a
fermion-fermion Feshbach resonance \cite{fsff}.  Demixing also appears
with the increase of trapping frequency for a smaller  $a_{12}$.
This later possibility to achieve demixing is more attractive
experimentally as it is easier to increase the trapping frequency than to
increase  $a_{12}$ near a Feshbach resonance by manipulating a background 
magnetic field \cite{fsff}. 
A proper treatment of DFFM
should be performed
using a fully antisymmetrized many-body Slater determinant wave function
\cite{yyy1} as in the case of atomic and molecular
scattering involving many electrons \cite{ps}. However, in
view of the success of the hydrodynamic model in other contexts
\cite{ska,fbs2,fds}
we do
not believe that the present study on mixing-demixing  in a DFFM
 to be so peculiar as to have no general validity.

 
\acknowledgments

The work is supported in part by the CNPq and FAPESP 
of Brazil.



\begin{thebibliography}{99}

 
\bibitem{exp1}B. DeMarco and D. S. Jin, Science {\bf 285}, 1703
(1999).
 
\bibitem{exp2} K. M. O'Hara, S. L. Hemmer, M. E. Gehm,  S. R. Granade,
and
J. E. Thomas, Science {\bf 298}, 2179 (2002).


 
\bibitem{exp3}F. Schreck, L. Khaykovich, K. L. Corwin, G. Ferrari, T.
Bourdel, J. Cubizolles, and C. Salomon, Phys. Rev. Lett. {\bf 87}, 080403
(2001); A. G. Truscott, K. E. Strecker, W. I. McAlexander, G. B.
Partridge, and R. G. Hulet, Science {\bf 291}, 2570 (2001).




 
 



 
  
 
\bibitem{exp4} Z. Hadzibabic, C. A. Stan, K. Dieckmann, S. Gupta, M. W.
Zwierlein, A. Gorlitz, and W. Ketterle, Phys. Rev. Lett. {\bf 88}, 160401
(2002); Z. Hadzibabic, S. Gupta, C. A. Stan, C. H. Schunck, M. W.
Zwierlein, K. Dieckmann, and W. Ketterle, {\it ibid.} {\bf 91}, 160401
(2003).
 







 


 
\bibitem{exp5}G. Modugno, G. Roati, F. Riboli, F. Ferlaino, R. J. Brecha,
and M. Inguscio, Science {\bf 297}, 2240 (2002).
 
 
 
\bibitem{exp5x} G. Roati,  F. Riboli, G. Modugno, and M. Inguscio,
Phys. Rev. Lett. {\bf 89}, 150403 (2002).
 





  
 
\bibitem{exp6}K. E. Strecker, G. B. Partridge, and R. G. Hulet,
Phys. Rev. Lett. {\bf 91}, 080406 (2003).
 
 
 
\bibitem{bongs} C. Ospelkaus, S. Ospelkaus, K. Sengstock, and
K. Bongs, Phys. Rev. Lett. {\bf 96}, 020401  (2006).    
 



\bibitem{zzz}M. Modugno,  F. Ferlaino,  F. Riboli,  G. Roati,  G.
Modugno,
and
M. Inguscio, 
 Phys. Rev.  A {\bf 68}, 043626 (2003).


 

\bibitem{ska}S. K. Adhikari, Phys. Rev. A {\bf 70}, 043617 (2004).

\bibitem{fbs2}S. K. Adhikari, Phys. Rev. A {\bf 72}, 053608 (2005).     



 
\bibitem{fds}S. K. Adhikari, J. Phys. B {\bf 38}, 3607 (2005).




\bibitem{md1} Y. Takeuchi and H. Mori,
Phys. Rev. A 72, 063617 (2005).

 
\bibitem{md2}  Z. Akdeniz, A. Minguzzi, P. Vignolo, and M.P. Tosi,
Phys. Lett. A {\bf 331}, 258 (2004);
P. Capuzzi, A. Minguzzi, and
M. P. Tosi,
 Phys. Rev. A  {\bf 68}, 033605
(2003).

 
\bibitem{yyy1} K. Molmer, Phys. Rev. Lett.
{\bf 80}, 1804 (1998).

 
\bibitem{bongs1}T. Karpiuk, M. Brewczyk, S. Ospelkaus-Schwarzer,
K. Bongs, M. Gajda, and K. Rz\c a\.zewski, Phys. Rev. Lett.     {\bf 93},
100401 (2004).   

\bibitem{demix1}R. Roth  and H. Feldmeier,   J. Phys. B {\bf 34}, 
4629 (2001).

\bibitem{demix2}A. Amoruso, I. Meccoli, A. Minguzzi and M. P. Tosi,
Eur. Phys. J. D {\bf 8},  361 (2000).


 
 
\bibitem{yyy}R. Roth, Phys. Rev. A {\bf 66}, 013614 (2002); R. Roth and
H. Feldmeier,  {\it ibid.}  {\bf 65}, 021603(R) (2002); T. Miyakawa,
T. Suzuki, and H. Yabu,  {\it ibid.}  {\bf 64}, 033611 (2001);
X.-J. Liu, M. Modugno, and H. Hu,
 {\it ibid.}  {\bf 68}, 053605 (2003).

 
 
 
 
 
 



 
\bibitem{capu}
P. Capuzzi, A. Minguzzi, and
M. P. Tosi,
 Phys. Rev. A
{\bf 67}, 053605 (2003).

\bibitem{rmp}F. Dalfovo, S. Giorgini, L. P. Pitaevskii, and
S. Stringari, Rev. Mod. Phys. {\bf 71}, 463 (1999).  

\bibitem{str} L. Vichi and  S. Stringari, Phys. Rev. A {\bf 60},
4734 (1999). 
 
\bibitem{sk1}S. K.  Adhikari  and P. Muruganandam, J. Phys. B
{\bf 35}, 2831 (2002); P.
 Muruganandam  and S. K.  Adhikari,  {\it ibid.}    {\bf 36}, 2501
(2003).


 
\bibitem{fsff} K. M. O'Hara, S. L. Hemmer, S. R. Granade, M. E. Gehm,  
J. E. Thomas, V. Venturi, E. Tiesinga, and C. J. Williams, 
Phys. Rev. A {\bf 66},
041401(R) (2002);
K. Dieckmann, C. A. Stan, S. Gupta, Z. Hadzibabic, C. H. Schunck, and
W. Ketterle, 
Phys. Rev. Lett. {\bf 89}, 203201
(2002); T. Loftus, C. A. Regal, C. Ticknor, J. L. Bohn, and
D. S. Jin,
{\it ibid.} {\bf 88}, 173201 (2002);  C. A. Regal, M. Greiner, and
D. S. Jin, {\it ibid.} {\bf 92},  083201 (2004).












 
 















 








 
 
 

 
 
 
 
 
 
 
 
 



 

 


\bibitem{ps}P. K.   Biswas and S. K.  Adhikari, J. Phys. B {\bf 33}, 1575
(2000); {\bf 31}, L737 (1998); {\bf 31}, 3147 (1998); {\bf 31}, L315
(1998).
 
 
 


 
\end{thebibliography}
\end{document}